\documentclass{article}
\usepackage{amssymb}
\usepackage{amsmath}

\begin{document}
\centerline{\Large \bf The introduction to the operator method}
\medskip
\centerline{\Large \bf for solving differential equations.}
\medskip
\centerline{\Large \bf First-order DE}

\vskip 2cm
\centerline{\sc Yu. N. Kosovtsov}
\medskip
\centerline{79060, 43-52, Nautchnaya Street, Lviv, Ukraine.}
\centerline{email: {\tt kosovtsov@escort.lviv.net}}
\vskip 1cm
\begin{abstract}
We introduce basic aspects of new operator method, which is very suitable for practical solving differential equations of various types. The main advantage of the method is revealed in opportunity to find compact exact operator solutions of the equations and then to transform them to more convenient form with help of developed family of operator identities. On example of non-linear first-order DEs we analyse analytical and algorithmical possibilities for solutions obtaining. Different forms of solutions for first-order DEs are given, including for some integro-differential equations and equations with variational derivatives. We describe new algorithms for direct computing the solutions with help of computer algebra system (CAS). We also discuss recipe for finding new solvability conditions, which allow to enlarge DE solving abilities of existent CAS.
\end{abstract}

\section{Introduction}

In recent years it becomes obvious a reanimation of intense interest to further development of exact methods for solving differential equations. The main reasons here are not only in expansion of science areas where differential equations play prominent role, but also in impetuous development of computer algebra systems (see \cite {MAH00}--\cite {Seiler}).

There are several universal approaches to solving DE. The most promising of them are Lie symmetry methods (see \cite {Olver} ) and advanced methods for the linear equations which are based on differential Galois theory \cite {Singer} with the first usable algorithm developed by Kovacic \cite {Kovacic}, \cite {Ulmer}. But in computer algebra very often underlying mathematics is far from the methods normally taught to mathematics students for the same problem \cite {MAH00}.

The aim of this contribution is to show on examples of first-order equations the possibilities for solving differential equations with new method based on Dyson's operator solution representation in form with time-ordered exponent. This method allows to find instantly the exact operator solution if we succeed in reformulating given problem to the first-order linear one. This solution is formal enough. The next step of the approach is directed to transformation the operator solution to combinations of some operators with relatively simple properties, which would permit to carry out all the operators and obtain as a result the solution of equation "by quadratures" or to more practical, calculable or usual (standard) expression.

The main advantages of the method are in its compactness, clarity and simpleness. Since we often can obtain some form of exact solution and due to the fact that it is much easier to study the "prepared frog". It is essential as well, that the operator method is algorithmic.

It is clear that there are some correlations between operator method and others common methods. But we believe that in introductory paper it is untimely to analyse such correlations.

We choose the first-order equation as a testing area for the operator method because it is most important differential equation in many branches of science and because main aspects of the method are quite easily generalized for other cases.

There are some known solvability conditions for the first-order equations (see for example \cite {Kamke}). But in the situation, when we are far from be able to solve any equation, the pursuit of new such conditions is very important.

To avoid induced complications we choose a nonrigourous manner of exposition inasmuch we always can verify obtained conclusions by substitution to initial equation. Analytical restrictions are obvious from context as a rule.

The paper is organized as follows. In Sec.2 we introduce basic ideas of the operator method. The applicability of the operator method for non-linear problems is discussed in Sec.3 on example of first-order non-linear ODE. The aim of this section is to introduce on particular case what we will do for the general cases. In Sec.4 we consider an idea of CAS solving algorithm, which is based on operator solution. We believe that this type of direct solving algorithms can play a noticeable part in future CAS developments. The main tools of operator method: the transformation identities are briefly described in Sec.5 and important properties of generalized shift operators are considered in Sec.6. In the Sec.7 we demonstrate algebraical way to simplify operator solutions. Such type of solutions can found more efficient direct CAS algorithms. The expression of the general solution of the equation through particular solution is considered in Sec.8. This section carries a methodical load too. Here we demonstrate the possibilities of operator method in obtaining solutions for more complicated equations. It is shown in Sec.9 that the solution of DE can be expressed in terms of arbitrary function. Among another consequences it brings some recipes for finding of solvability conditions that allow to enlarge DE solving abilities of existent CAS.

\section{Base of the operator method}

The main point of the operator method is based on solution of the \emph{first-order linear} equation obtained by Dyson in the form with time-ordered exponent.

 For the \emph{first-order} equation with respect to $t$  (in ordinary or partial derivatives) for $u(t,\vec{\rho})$ :
\begin{equation}
\frac{\partial u(t,\vec{\rho})}{\partial t} =\hat{L}(t,\vec{\rho})u(t,\vec{\rho})+f(t,\vec{\rho})
\label{base}
\end{equation}
with the initial condition
\begin{equation}
u(a,\vec{\rho})=v(\vec{\rho}),
\label{basein}
\end{equation}
where $\hat{L}$  is a \emph{linear operator} (independent variable $\vec{\rho}$\,  is one dimensional or many dimensional),  the solution is given by $(t\geq a)$
\begin{equation}
u(t,\vec{\rho})=\hat{T}\, \exp\{\int_a^t d\tau\,\hat{L}(\tau,\vec{\rho})\}v(\vec{\rho})+\int_a^t d\xi\,\hat{T}\, \exp\{\int_\xi^t d\zeta\,\hat{L}(\zeta,\vec{\rho})\}f(\xi,\vec{\rho})
\label{basesolut} \ .
\end{equation}
Here the operator  $\hat{T}$\, is a "chronological" (ordering) Dyson's operator, representing a product, ordered (in order of unincrease) along the co-ordinate $t$ , of the operators, which appear in the integrands in the expansion (see below) of the exponent in an operator series:
\begin{align}
\hat{T}\,[\hat{L}(\tau_1,\vec{\rho})\dots \hat{L}(\tau_m,\vec{\rho})]= \hat{L}(\tau_{\alpha_1},& \vec{\rho})\hat{L}(\tau_{\alpha_2},\vec{\rho})\dots \hat{L}(\tau_{\alpha_m},\vec{\rho})\ .\notag \\ & \tau_{\alpha_1}\geq \tau_{\alpha_2}\geq \dots \geq \tau_{\alpha_m}\notag
\end{align}
This chronological operator has to be introduced because the operators in the integrands with different values of the variable $\tau$  may not commute. The solution of the problem (\ref{base}), (\ref{basein}) for $t<a$   is similar to (\ref{basesolut}), but instead of the operator $\hat{T}\,$  it is necessary to place the operator $\hat{T}_0\,$ , which represents a product, ordered in reverse order.

We can define chronological operator exponent by another way, e.g., through its differential properties or by way of expansion in an operator series. We can find that
\[\hat{T}\, \exp\{\int_a^t d\tau\,\hat{L}(\tau,\vec{\rho})\}=\hat{1}+\int_a^t d\tau\,\hat{L}(\tau,\vec{\rho})\,\hat{T}\, \exp\{\int_a^\tau d\xi\,\hat{L}(\xi,\vec{\rho})\} \ ,
\]
were $\hat{1}$  is the identity operator. Father iterations lead to
\begin{align}
\hat{T}\, \exp\{& \int_a^t d\tau\, \hat{L}(\tau,\vec{\rho})\}= \hat{1}+\int_a^t d\tau\,\hat{L}(\tau,\vec{\rho})+\dots\notag \\ & \dots +\int_a^t d\tau_1\int_a^{\tau_1} d\tau_2\dots \int_a^{\tau_{n-1}} d\tau_n\,\hat{L}(\tau_1,\vec{\rho})\,\hat{L}(\tau_2,\vec{\rho})\dots \hat{L}(\tau_n,\vec{\rho})+ \dots \label{opseries}
\end{align}

The most important here is the fact that (\ref{basesolut}) is really the exact solution of the equation (\ref{base}) with the initial condition (\ref{basein}). One can very easily verify it by means of differentiation of (\ref{basesolut}) with respect to $t$ .

It is obvious that
\[
\hat{T}\, \exp\{\int_a^t d\tau\, \hat{L}(\tau,\vec{\rho})\}\,\,\hat{T}_0\, \exp\{\int_t^a d\tau\, \hat{L}(\tau,\vec{\rho})\}=\,\hat{1}
\]
and, e.g., when $a\leq t\leq b$
\[
\hat{T}\, \exp\{\int_a^b d\tau\, \hat{L}(\tau,\vec{\rho})\}\,\,\hat{T}_0\, \exp\{\int_b^t d\tau\, \hat{L}(\tau,\vec{\rho})\}=\hat{T}\, \exp\{\int_a^t d\tau\, \hat{L}(\tau,\vec{\rho})\}.
\]
It must be stressed here that we do not involve any premises about nature of $u(t,\vec{\rho})$ , it can be both scalar and vectorial function, and it can depend on any number of independent arguments. For operator $\hat{L}$  (and its "filler") as well - the only fundamental requirement - it must be a linear operator. Very often non-linear problem is reduced to linear differential equation. If we succeed in reformulating given problem to first-order linear one, we then instantly can find its solution in operator form.

At first sight the chronological operator exponent is a very specific case of linear operators. The following proposition asserts the opposite. Let we have arbitrary but \emph {linear differentiable reversible} operator $\hat{A}(t)$ . From obvious identity
\[\frac{d \hat{A}(t)}{dt}=[\frac{d \hat{A}(t)}{dt} \hat{A}^{-1}(t)]\hat{A}(t)\ , \qquad \hat{A}(t)|_{t=a}=\hat{A}(a)
\]
we obtain
\[ \hat{A}(t)=\hat{T}\, \exp\{\int_a^t d\tau\,[\frac{d \hat{A}(\tau)}{d\tau} \hat{A}^{-1}(\tau)]\}\hat{A}(a) \ ,
\]
so the chronological operator exponent can represent \emph {any} linear differentiable reversible operator, only this form is exactly adjusted for solving defined DE.
For any operator of such type there is corresponding linear DE.
We can stress here that to solve a DE in explicit form we have to express explicitly the action of corresponding linear operator. It is the challenge: far from all linear operators can be specified by explicit action.

In present paper for simplicity we restrict consideration by first-order DE's.

\section{Operator solutions for first-order non-linear ODE}

First-order differential equation of arbitrary form
\begin{equation}
\frac{du(t)}{dt}=f(t,u(t))
\label{ode1}
\end{equation}
on account of non-linearity is not suited immediately for solving by operator method. Nevertheless there are many possibilities to convert the problem to linear one. Most of them are concerned with introduction of spaces of larger dimensions. Here we demonstrate some of them, which are almost generally applicable.

If we introduce a new function
\[S(t,\omega)=e^{\omega u(t)} \ , \]
from (\ref{ode1}) one can easily derive the following equivalent to (\ref{ode1})\, first-order (with respect to $t$) linear DE for $S(t,\omega)$
\begin{equation}
\frac{\partial S(t,\omega)}{\partial t}=\omega f(t,\frac{\partial}{\partial \omega})S(t,\omega)\ , \qquad S(a,\omega)=e^{\omega u(a)},
\label{Lode1}
\end{equation}
which formal solution follows immediately from (\ref{basesolut})\,in the form
\begin{equation}
S(t,\omega)=\hat{T}\, \exp\{\int_a^t d\tau\,\omega f(\tau,\frac{\partial }{\partial \omega})\}e^{\omega u(a)}.
\label{Lode1sol}
\end{equation}
The solution of equation (\ref{ode1})\,is obvious from (\ref{Lode1sol}):
\[u(t)=\frac{\partial S(t,\omega)}{\partial \omega}|_{\omega=0}.
\]

We have to note here that when we look for general or any particular solutions we can utilize indefinite integrals in operator solutions.

    The operator solutions of type (\ref{basesolut}), (\ref{Lode1sol}) are formal enough. Meantime, solutions satisfy the following requirements to be the exact solution:

a). Unknown is on the left side. The right side consists of only and only knowns (variables, coefficients, functions, operators and etc.).

b). Direct substitution of the solution reduces the equation to an identity.

The theorem for uniqueness of solution of differential equation shows that if we have an exact solution of a problem in several forms, all the forms can be transformed to each other. So the main problem that remains in this approach is how to transform of the operator solution to that form which can be considered as usable.

The most part of this paper is devoted to hunting such ways.

\section{An idea of solving algorithm}

It would be rather well if we can find a universal algorithm, which would solve any DE. The most attractive feature of the operator method is unusual easiness in obtaining of formal (operator) solutions of DE's. Can we find on this way a general algorithm to obtain a standard (non-operator) solutions of DEs?

Let us consider \emph {an idea} of such algorithm on special example of solution of equation for unknown function $u(t)$:
\begin{equation}
\frac{du(t)}{dt}=u^2(t)+\frac{da(t)}{dt}-a^2(t)\ ,
\label{exode}
\end{equation}
where $a(t)$ is some known function. Such type of equation is very suitable for \emph {testing} algorithms. First of all since its particular solution is obviously known: $u(t)=a(t)$. In the second place since existent CA systems \emph {do not solve} at present most cases of such type of equations (with specific expressions for $a(t)$).

The operator solution of equation (\ref{exode}) in form (\ref{Lode1sol})\,is
\begin{equation}
u(t)=[\frac{\partial }{\partial \omega}\,\hat{T}\, \exp\{\int dt\,\omega [\frac{\partial^2 }{\partial \omega^2}+\frac{da(t)}{dt}-a^2(t)]\}e^{\omega c}]_{\omega=0},
\label{solexode}
\end{equation}
were $c$ is an arbitrary constant (farther on in this section for simplicity let $c=0$, i.e. we will seek here only a particular solution of equation). If we represent the chronological exponent as the series (\ref{opseries}):

\[ u(t)=[\frac{\partial }{\partial \omega}\,\sum_{i=1}^{\infty}S_i(\omega,t)]_{\omega=0}, \]
terms of series obviously obey the following recurrence relations
\[S_1(\omega,t)=1 \ , \qquad S_i(\omega,t)=\int dt\,\omega [\frac{\partial^2}{\partial \omega^2}+\frac{da(t)}{dt}-a^2(t)]S_{i-1}(\omega,t). \]
Now let us use the following trick. If we compute the sequence of
\[ u_n(t)=[\frac{\partial }{\partial \omega}\,\sum_{i=1}^n S_i(\omega,t)]_{\omega=0}\ ,\]
we find
\begin{align}
u_1(t)&=a(t)-\int dt\,a^2(t) \ ,  \notag\\
u_3(t)&=a(t)-2\int dt\,a(t)\int dt\,a^2(t)+\int dt\,[\int dt\,a^2(t)]^2 \ ,  \notag\\
u_5(t)&=a(t)-3\int dt[\int dt\,a^2(t)]^2+6\int dt\int dt\,a(t)[\int dt\,a^2(t)]^2+ \dots \notag
\end{align}
If compare now options in sequence of $u_n(t)$ we can find that from some $n_0$ an unaltered (survived) part in $u_n(t)$ become apparent (the $a(t)$ here). It is reasonable that we can suppose that it is just the sought solution. On solution of DEs we always have the opportunity to verify it. If we get affirmative answer, so it is the end of procedure, else we proceed with larger $n$.

We have prepared a computer algebra prototype of some variants of the algorithm using the Maple system. The algorithm manages, e.g. the following DE's (which we choose mostly by chance):
\begin{align}
DE1 := \frac{du(t)}{dt} = & u^2(t)+\frac { 1}{t^{138}}+41552420\, t^{90}+4410 \, t^{89}+\frac{7225}{t^{136}}-13430\, t^{23}+\notag\\&+8330 \,t^{22}+6241 \,t^{182}-7742 \, t^{181}-2401 \,t^{180}+1,\notag\\
& Ans1 := u(t) = -\frac{85}{t^{68}}+79\,t^{91}+49\,t^{90}; \notag
\end{align}
\begin{align}
DE2 := \frac{du(t)}{dt} = & u^2(t)-p\,m\,\cos(t)\,\sin^{(m-1)}(t)+p^2\,\sin^{2m}(t),\notag\\
& Ans2 := u(t) = p\,\sin^m(t);\notag
\end{align}
\begin{align}
DE3 := \frac{du(t)}{dt} = & u^2(t)+p\,t^{m-1}m\,e^{nt}+p\,t^m\,n\,e^{nt}+q\,l\,\frac{\ln^{l-1}(t)}{t}-\notag\\&-t^{2m}\,p^2\,e^{2nt}-2\,t^m\,p\,e^{nt}\,q\,\ln^l(t)-q^2\,\ln^{2l}(t),\notag\\
& Ans3 := u(t) = p\,t^m\,e^{nt}+q\,\ln^l(t).\notag
\end{align}

Let us make some remarks here. For specific equation we can transform operator solution to improve, e.g. rate of convergence. In the cases under consideration in this section it is very useful with help of operator identities (see below \,(\ref{baseId1}) (\ref{baseId2})) to modify the operator solution to the following form:
\begin{align}
u(t)=&[\frac{\partial }{\partial \omega}\,\exp\{\int dt\,\omega [\frac{da(t)}{dt}-a^2(t)]\}\,\hat{T}\, \exp\{\int dt\,\omega \times \notag\\&\times\exp\{-\int dt\,\omega [\frac{da(t)}{dt}-a^2(t)]\times \notag\\&\times\frac{\partial^2 }{\partial \omega^2}\exp\{\int dt\,\omega [\frac{da(t)}{dt}-a^2(t)]\}e^{\omega C}]_{\omega=0}.
\notag
\end{align}

Experimentation with prototype of the algorithm leads us to conclusion that the \emph { algorithm is workable} and can solve (at least in sense of particular solutions) a set of equations which Maple cannot solve now. It is really the \emph {solving} algorithm as against from algorithms based on certain solvability conditions or mapping into separable equations.

Central failure of this algorithm consists in the fact that it \emph {heavily relies on basic CAS procedures and requires noticeable resources}.

It is clear that we can try to find \emph {some} particular solutions (under different values of $c$) or even \emph {general} solution of DE by the same way and exploit any form of operator solutions. However the summation of functional series with several variables is for existent CAS substantially more resources consumimg task.

In next sections we examine some ways for solution transformations, which would allow to simplify procedure of solutions calculation.

\section{Basic transformation identities}

Our aim is to reduce the equation solution from operator form (\ref{basesolut}) to more practical, calculable or usual (standard) expression. To do so it is necessary to find ways to evaluate chronological operator exponents or convert it to combinations of some operators with relatively simple properties, which would permit to carry out all the operators and obtain as a result the solution of equation "by quadratures".

One of the main tool kits in the operator method is a family of identities for chronological operator exponents for solutions transformations.

At present time the most-used method to "disentangle" of operator exponents is well-known Campbell-Baker-Hausdorff (CBH) expansion, which for chronological exponents has the following version:
\begin{align}
\hat{T}_0\,& \exp\{-\int_a^\tau d\xi\,\hat{A}(\xi)\}\,\hat{B}(\tau)\,\hat{T}\, \exp\{\int_a^\tau d\xi\,\hat{A}(\xi)=\hat{B}(\tau)+\int_a^\tau d \tau_1\,[\hat{B}(\tau),\hat{A}(\tau_1)]+\dots \notag \\ & \dots +\int_a^\tau d\tau_1\int_a^{\tau_1} d\tau_2\dots \int_a^{\tau_{n-1}} d \tau_n\,[[\dots[[\hat{B}(\tau),\hat{A}(\tau_1)],\hat{A}(\tau_2)]\dots ],\hat{A}(\tau_n)] +\notag  \\& + \int_a^\tau d \tau_1\int_a^{\tau_1} d\tau_2\dots \int_a^{\tau_n} d\tau_{n+1}\,\hat{T}_0\, \exp\{-\int_a^{\tau_{n+1}} d \xi\,\hat{A}(\xi)\}\,\times \notag \\ &\times[[\dots[[\hat{B}(\tau),\hat{A}(\tau_1)],\hat{A}(\tau_2)]\dots ],\hat{A}(\tau_{n+1})]\,\hat{T}\, \exp\{\int_a^{\tau_{n+1}} d \xi\,\hat{A}(\xi)\}, \notag
\end{align}
where $[\hat{B}(\tau),\hat{A}(\tau_1)]=\hat{B}(\tau)\hat{A}(\tau_1)-\hat{A}(\tau_1)\hat{B}(\tau)$  is the standard notation of operators commutator.

For purpose of this paper the following related identities are of fundamental importance (all operators are supposed to be linear):

\begin{align}
& \hat{T}\, \exp\{\int_a^t d\tau\,\hat{A}(\tau)\}\, \hat{T}\, \exp\{\int_a^t d\tau\,\hat{B}(\tau)\}= \notag \\ & = \hat{T}\, \exp\{\int_a^t d\tau\,[\hat{A}(\tau)+\hat{T}\, \exp\{\int_a^\tau d\xi\,\hat{A}(\xi)\}\,\hat{B}(\tau)\,\hat{T}_0\, \exp\{-\int_a^\tau d\xi\,\hat{A}(\xi
)\}]\}
\label{baseId1}
\end{align}
or in more general form
\begin{align}
\hat{A}(t)\,&\hat{T}\, \exp\{\int_a^t d\tau\,\hat{B}(\tau)\}=\notag \\&=\hat{T}\, \exp\{\int_a^t d\tau\,[(\frac{\partial \hat{A}(\tau)}{\partial \tau})\hat{A}^{-1}(\tau)+\hat{A}(\tau)\,\hat{B}(\tau)\,\hat{A}^{-1}(\tau)\,\}\hat{A}(a)
\label{baseId11}
\end{align}
end
\begin{align}
\hat{T}\, \exp\{& \int_a^t d\tau\,[\hat{A}(\tau)+\hat{B}(\tau)]\}=\hat{T}\, \exp\{\int_a^t d\tau\,\hat{A}(\tau)\}\times \notag \\ & \times \hat{T}\, \exp\{\int_a^t d\tau\,\hat{T}_0\, \exp\{-\int_a^\tau d\xi\,\hat{A}(\xi)\}\,\hat{B}(\tau)\,\hat{T}\, \exp\{\int_a^\tau d\xi\,\hat{A}(\xi
)\}\}.
\label{baseId2}
\end{align}
Correctness of these identities is easily verified by differentiations (the left and right sides satisfy the same differential equations with the same initial conditions).

From identity (\ref {baseId2})\,follows the identity for chronological exponents with reverse ordering:

\begin{align}
\hat{T}_0\, \exp\{& \int_a^t d\tau\,[\hat{A}(\tau)+\hat{B}(\tau)]\}=\notag \\ & =\hat{T}_0\, \exp\{\int_a^t d\tau\,\hat{T}_0\, \exp\{\int_a^\tau d\xi\,\hat{A}(\xi)\}\,\hat{B}(\tau)\,\hat{T}\,\exp\{-\int_a^\tau d\xi\,\hat{A}(\xi
)\}\}\times \notag \\ & \times \hat{T}_0\, \exp\{\int_a^t d\tau\,\hat{A}(\tau)\}.
\label{baseId3}
\end{align}

We will need later an variant of identity of type (\ref {baseId11}), when we do not demand from stationary operator $\hat{D}$ to be reversible:
\begin{align}
\hat{D}\,\hat{T}\, \exp\{\int_a^t d\tau\,\hat{B}(\tau)\,\hat{D}\}=\hat{T}\, \exp\{\int_a^t d\tau\,\hat{D}\,\hat{B}(\tau)\,\}\hat{D}.
\label{baseId12}
\end{align}

The identity (\ref{baseId1})\, imparts group properties to chronological operators. We do not evolve here this interesting matter.

\section{Generalized shift operators}

To transform formal solutions to ordinary expressions we have to express explicitly the action of corresponding linear operators. Every solved equation gives us an example of definite action of given linear operator. So the list of "good" operators is nonempty. Our goal here is to study the properties of certain class of operators that we will name as generalized shift operators.

    The Taylor expansion for sufficiently arbitrary function $\Phi(x)$ can be expressed as
\[ \Phi(x+\alpha)=\sum_{k=0}^{\infty} \frac{1}{k!}\alpha^k \frac{d^k}{dx^k} \Phi(x)= \exp{\{\alpha \frac{d}{dx}\}}\Phi(x)\ .
\]
Here $\alpha$  can be either constant or function of arguments, which are different from $x$ \,(in the second case $\frac{d}{dx}$ is replaced by $\frac{\partial}{\partial x}$ . If we read this expression from the right side, we can obtain the result of action of exponential form of \emph {shift operator} $\exp{\{\alpha \frac{d}{dx}\}}$  on function $\Phi(x)$ :
\[ \exp{\{\alpha \frac{d}{dx}\}}\Phi(x) = \Phi(\exp{\{\alpha \frac{d}{dx}\}}x)= \Phi(x+\alpha) \]
or
\begin{equation} \exp{\{\alpha(t) \frac{\partial}{\partial x}\}}\Phi(x) = = \Phi(\exp{\{\alpha(t) \frac{\partial}{\partial x}\}}x)=\Phi(x+\alpha(t)). \label{Shift1}
\end{equation}

Let us consider shift operators with new variable $y=\psi(x)$ . It is obvious, that
\[ \exp{\{\alpha(t) \frac{\partial}{\partial y}\}}\Phi(y) = \exp{\{\alpha(t) \frac{\partial}{\partial \psi(x)}\}}\Phi(\psi(x))=\]
\[=\exp{\{\alpha(t)\frac{1}{\psi(x)'} \frac{\partial}{\partial x}\}}\Phi(\psi(x))=\Phi(\psi(x)+\alpha(t)). \]
When $\psi(x)$ is fixed function, the set of shift operators with different parameters $\alpha(t)$  forms Abelian group. Shift operators with different $\psi(x)$ do not commute.

It is quite obvious that all examined shift operators correspond to DEs with separable variables.

Let us consider now one important property of the operators of the following type

\[\hat{T} \exp\{\int_a^t d\tau\,f(\tau,x)\frac{\partial }{\partial x}\},\]
which we will call as generalized shift operators.

Let $g(t,x)$ and $f(t,x)$ are ordinary functions (linear operators of multiplication by value of function). Since the commutator
\[ [g(t,x),f(\tau,x)\frac{\partial}{\partial x}]=-f(\tau,x)(\frac{\partial g(t,x)}{\partial x}) \]
is ordinary function, so from CBH - expansion we can conclude that following expression

\[ G(t,x)=\hat{T} \exp\{\int_a^t d\tau\,f(\tau,x)\frac{\partial }{\partial x}\}\,g(t,x)\,\hat{T}_0 \exp\{-\int_a^t d\tau\,f(\tau,x)\frac{\partial }{\partial x}\}\]
is ordinary function too and we can compute it simply as
\begin{equation}
G(t,x)=\hat{T} \exp\{\int_a^t d\tau\,f(\tau,x)\frac{\partial }{\partial x}\}\,g(t,x).
\label{G}
\end{equation}

We find that
\begin{align}
\hat{T} & \exp \{\int_a^t d\tau\,f(\tau,x)\frac{\partial }{\partial x}\}\,g_1(t,x)\,g_2(t,x)=\notag \\ & =\hat{T} \exp\{\int_a^t d\tau\,f(\tau,x)\frac{\partial }{\partial x}\}\,g_1(t,x)\,\hat{T}_0 \exp\{-\int_a^t d\tau\,f(\tau,x)\frac{\partial }{\partial x}\}\times\notag \\ &\times \hat{T} \exp\{\int_a^t d\tau\,f(\tau,x)\frac{\partial }{\partial x}\}\,g_2(t,x)=\notag \\ & =[\hat{T} \exp\{\int_a^t d\tau\,f(\tau,x)\frac{\partial }{\partial x}\}\,g_1(t,x)][\hat{T} \exp\{\int_a^t d\tau\,f(\tau,x)\frac{\partial }{\partial x}\}\,g_2(t,x)]=\notag \\ & =G_1(t,x)\,G_2(t,x).
\label{G2}
\end{align}
If now expand $g(t,x)$ into the Taylor power series with respect to $x$, use (\ref{G2})\ and then collect series backwards, we obtain analogous to (\ref{Shift1})\, very important property (homomorphy):

\begin{equation}
G(t,x)=\hat{T} \exp\{\int_a^t d\tau\,f(\tau,x)\frac{\partial }{\partial x}\}\,g(t,x)=g(t,\hat{T} \exp\{\int_a^t d\tau\,f(\tau,x)\frac{\partial }{\partial x}\}\,x).
\label{Homomorphy}
\end{equation}
Now we are ready to consider another variants of operator solutions for equation (\ref{ode1}) and farther modifications of solving algorithm.

\section{Solution for first-order ODE in form of generalized shift operator}

Let us start from operator solution of equation (\ref{ode1}) in form (\ref{Lode1sol}) but let us rewrite it for convenience in following notation
\[S(t,c,\omega)=\hat{T}\, \exp\{\int_a^t d\tau\,\omega f(\tau,\frac{\partial }{\partial \omega})\}\,e^{\omega c},\]
where $c=u(t)|_{t=a}$. With help of identity (\ref{baseId11}) we find
\[S(t,c,\omega)=e^{\omega c}\,\hat{T}\, \exp\{\int_a^t d\tau\,\omega \,e^{-\omega c} f(\tau,\frac{\partial }{\partial \omega})e^{\omega c}\}\cdot1.\]
CHB-expantion gives
\begin{equation}
e^{-\omega c} f(\tau,\frac{\partial }{\partial \omega})e^{\omega c}=f(\tau,\frac{\partial }{\partial \omega})+c[f(\tau,\frac{\partial }{\partial \omega}),\omega]+\frac{c^2}{2}[[f(\tau,\frac{\partial }{\partial \omega}),\omega],\omega]+\dots
\label{Expr}
\end{equation}
Since it is easily proved by mathematical induction that
\[[\frac{\partial^m }{\partial \omega^m},\omega]=m\frac{\partial^{m-1} }{\partial \omega^{m-1}},\]
then from expantion of $f(x)$ into power series and backwards summation it follows that
\[[f(\frac{\partial }{\partial \omega}),\omega]=f'(\frac{\partial }{\partial \omega})\qquad (f'(x)=\frac{\partial f(x) }{\partial x}),\]
therefore we can conclude with taking into account (\ref{Expr}) and property of the shift operator (\ref{Shift1}) that
\[e^{-\omega c} f(\tau,\frac{\partial }{\partial \omega})e^{\omega c}=f(\tau,\frac{\partial }{\partial \omega})+c\,f'(\tau,\frac{\partial }{\partial \omega})+\frac{c^2}{2}\,f''(\tau,\frac{\partial }{\partial \omega})+\dots=\]

\[=f(\tau,c+\frac{\partial }{\partial \omega})=\exp\{\frac{\partial }{\partial \omega}\frac{\partial }{\partial c}\} f(\tau,c)  \exp\{-\frac{\partial }{\partial \omega}\frac{\partial }{\partial c}\}.\]
Then
\begin{align}
S(t,c,\omega)& =e^{\omega c}\exp\{\frac{\partial }{\partial \omega}\frac{\partial }{\partial c}\}\hat{T}\, \exp\{\int_a^t d\tau\,(\omega-\frac{\partial }{\partial c}) f(\tau,c)\}\,\exp\{-\frac{\partial }{\partial \omega}\frac{\partial }{\partial c}\}\cdot 1 =\notag \\ & = e^{\omega c}\exp\{\frac{\partial }{\partial \omega}\frac{\partial }{\partial c}\}\hat{T}\, \exp\{\int_a^t d\tau\,[\omega f(\tau,c)-\frac{\partial f(\tau,c)}{\partial c} -f(\tau,c)\frac{\partial}{\partial c}] \} \cdot 1. \notag
\end{align}
By expanding the chronological exponent with help of identity (\ref{baseId2}) and properties (\ref{G}), (\ref{G2})\, of generalized shift operators we obtain that
\begin{align}
S( t,c,\omega)& =e^{\omega c}\exp\{\frac{\partial }{\partial \omega}\frac{\partial }{\partial c}\}\hat{T}\, \exp\{-\int_a^t d\tau\,f(\tau,c)\frac{\partial}{\partial c}\} \times \notag \\ & \times \exp\{-\int_a^t d\tau\,g(\tau,c)\} \exp\{\omega\int_a^t d\tau\,G(\tau,c)\},
\end{align}
where
\[ g(\tau,c)=\hat{T}_0\, \exp\{\int_a^\tau d\xi\,f(\xi,c)\frac{\partial}{\partial c}\}\,\frac{\partial f(\tau,c)}{\partial c} \]
and
\[ G(\tau,c)=\hat{T}_0\, \exp\{\int_a^\tau d\xi\,f(\xi,c)\frac{\partial}{\partial c}\}\,f(\tau,c). \]
If we differentiate the last expression with respect to $\omega$ we find that
\begin{align}
\frac{\partial S(t,c,\omega) }{\partial \omega}&=c\,S(t,c,\omega)+e^{\omega c}\exp\{\frac{\partial }{\partial \omega}\frac{\partial }{\partial c}\}\hat{T}\, \exp\{-\int_a^t d\tau\,f(\tau,c)\frac{\partial}{\partial c}\}\times\notag \\ &\times  \exp\{-\int_a^t d\tau\,g(\tau,c)\} \exp\{\omega\int_a^t d\tau\,G(\tau,c)\}\,\int_a^t d\tau\,G(\tau,c).
\notag
\end{align}
If in the second item to recover initial operator form we receive
\[\frac{\partial S(t,c,\omega) }{\partial \omega}=c\,S(t,c,\omega)+ S(t,c,\omega)\,\int_a^t d\tau\,G(\tau,c), \]
inasmuch as $S(t,c,\omega)|_{\omega=0}\equiv 1$, then
\[ u(t,c)= \frac{\partial S(t,c,\omega) }{\partial \omega}|_{\omega=0}=c+ \int_a^t d\tau\,G(\tau,c),\]
whence it follows that
\[ u(t,c)=c+ \int_a^t d\tau\,\hat{T}_0\, \exp\{\int_a^\tau d\xi\,f(\xi,c)\frac{\partial}{\partial c}\}\,f(\tau,c)\]
or finally
\begin{equation}
u(t,c)=\hat{T}_0\, \exp\{\int_a^t d\tau\,f(\tau,c)\,\frac{\partial}{\partial c}\}\,c.
\label{Lode2sol}
\end{equation}
All above consideration is nonrigourous enough. So it is very important to verify our conclusion by direct substitution of obtained solution into equation (\ref{ode1}). After differentiation of (\ref{Lode2sol}) and with property (\ref{Homomorphy}) we can be sure that (\ref{Lode2sol}) is really the solution of the equation (\ref{ode1}).

So we have obtained solution of first-order equation in more simple operator form. We can prepare now a prototype of solving algorithm based on solution (\ref{Lode2sol}) by the way similar to considered in Section 4. It would be easily found that in this case algorithm can handle more complicated equations and with more efficiency. But it remains here (in sense of present CAS abilities) the complication in obtaining general solutions too. In following Section we consider possibilities to obtain general solution out of known particular solution.

Let us finish this Section with brief remark about interconnection between first-order ODE and linear first-order PDE.

If we introduce the function
\begin{equation}
z(t,c)=\hat{T}\, \exp\{-\int_a^t d\tau\,f(\tau,c)\frac{\partial}{\partial c}\}\,c
\label{Pdesol}
\end{equation}
it is obvious that it satisfies equation
\[\frac{\partial z(t,c)}{\partial t}+f(t,c)\,\frac{\partial z(t,c)}{\partial c}=0.\]
If we act now by operator
\[ \hat{T}_0\, \exp\{\int_a^t d\tau\,f(\tau,c)\frac{\partial}{\partial c}\} \]
on both sides of (\ref{Pdesol}), we get
\[ z(t,u(t,c))=c  \]
-- the well-known equation expressing interconnection between solutions of first-order ODE and linear first-order PDE.

Besides that that we \emph {obtain} (operator) solutions of the equations, a nuance here consists in fact that conventional derivation of this interconnection is carried out by means of geometrical considerations
(method of characteristics). We use only algebraic transformations and this approach is applicable for more complicated problems. We have the chance to ascertain it in the next Section.

\section{General solution in terms of one particular solution}

If we are able to find some particular solutions of equation (\ref{ode1}), then arises the question: can we express the general solution of the equation through these particular solutions? One of the possible answers is the famous solution of this problem by Lie \cite {Lie}. In terms of operator method Lie's result is: the general solution can be expressed by relatively simple function of set of particular solutions iff operator solution (\ref{Lode1sol}) can be expressed as a product of simple shift operators of type (\ref{Shift1}). On the other hand it is well-known that, e.g., if we know only one particular solution for Riccati equation the general solution can be found through quadratures.

Let us analyse this question from operator method point of view.

From solution (\ref{Lode2sol}) and with help of identities (\ref{baseId11}), (\ref{baseId12}) we find that
\begin{align}
\frac{\partial u(t,c)}{\partial c} &=\hat{T}_0\, \exp\{\int_a^t d\tau\,\frac{\partial}{\partial c}\,f(\tau,c)\}\cdot 1 =\notag \\ & =\hat{T}_0\, \exp\{\int_a^t d\tau\,[\,\frac{\partial f(\tau,c)}{\partial c}\,f(\tau,c)+f(\tau,c)\frac{\partial}{\partial c}]\,\}\cdot 1=\notag \\ & =\exp\{\int_a^t d\tau\, \hat{T}_0\, \exp\{\int_a^\tau d \xi\,f(\xi,c)\,\frac{\partial}{\partial c}\}  \frac{\partial f(\tau,c)}{\partial c}\}.\notag
\end{align}
If denote
\[\frac{\partial f(\tau,c)}{\partial c}=\Phi(\tau,c),\]
then on account of (\ref{Lode2sol}) we can find that $u(t,c)$ satisfies the following integro-differential equation:
\begin{equation}
\frac{\partial u(t,c)}{\partial c}=\exp\{\int_a^t d\tau\,\Phi(\tau,u(\tau,c))\,\}.
\label{Id}
\end{equation}
Solution of such type of equations is the challenge.

Let us introduce the following functional
\[S[c;\omega(\xi)]= \exp\{\int_{-\infty}^\infty d \xi\,\omega(\xi)\,u(\xi,c) \}.\]
Since
\[\frac{\partial S[c;\omega(\xi)]}{\partial c}=S[c;\omega(\xi)]\int_{-\infty}^\infty d \xi\,\omega(\xi)\,\frac{\partial u(\xi,c)}{\partial c},\]
then to the integro-differential equation (\ref{Id}) corresponds linear equation for $S[c,\omega(\xi)]$ with variational derivatives
\[\frac{\partial S[c;\omega(\xi)]}{\partial c}= \int_{-\infty}^\infty d \xi\,\omega(\xi)\, \exp\{\int_a^\xi d\tau\,\Phi(\tau,\frac{\delta }{\delta \omega(\tau)})\}S[c;\omega(\xi)]\]
with initial condition
\[ S[c;\omega(\xi)]|_{c=c_0}= \exp\{\int_{-\infty}^\infty d \xi\,\omega(\xi)\,u_0(\xi)\}, \qquad  u_0(t)=u(t,c)|_{c=c_0} .\]

Its operator solution is
\begin{align}
&S[c;\omega(\xi)] =\notag \\ & =\hat{T}_c\,\exp\{\int_{c_0}^c d\sigma\int_{-\infty}^\infty d \xi\,\omega(\xi)\,\exp \{\int_a^\xi d\tau\,\Phi(\tau,\frac{\delta }{\delta \omega(\tau)})\}\,\}\exp\{\int_{-\infty}^\infty d \xi\,\omega(\xi)\,u_0(\xi)\}.\notag
\end{align}

It is obvious that solution in this form is very intricate. So we have to simplify it. If we repeat the chain of transformations almost literally like that in preceding Section, we obtain the solution in more simple form
\begin{equation}
u(t,c)=\,\exp\{(c-c_0)\int_{-\infty}^\infty d \xi\,\exp \{\int_a^\xi d\tau\,\Phi(\tau,u_0(\tau))\,\}\frac{\delta }{\delta u_0(\xi)}\}\,\,u_0(t).
\label{Idsol}
\end{equation}
Now we have to examine that (\ref{Idsol}) is really the solution of equations (\ref{Id}) and (\ref{ode1}). The verification is not difficult matter here.

So we conclude that if given first-order ODE has general solution it can be expressed in terms of one particular solution of the equation.
Certainly we can find the well-known result for Riccati equation from (\ref{Idsol}).

The expression (\ref{Idsol}) represents the algorithm for computing the general solution from known particular solution. We have not tried to experiment with this algorithm yet.

\section{General solution in terms of arbitrary function and solvability conditions}

Let now $f(t,c)=g(t,c)+h(t,c)$, then we can find the following form of general solution of equation (\ref{ode1})
\begin{align}
u(t,c)&=\hat{T}_0\, \exp\{\int_a^t d\tau\,f(\tau,c)\,\frac{\partial}{\partial c}\}\,c\,=\hat{T}_0\, \exp\{\int_a^t d\tau\,[g(\tau,c)+h(\tau,c)]\,\frac{\partial}{\partial c}\}\,c=\notag \\ & =\hat{T}_0\, \exp\{\int_a^t d\tau\, [\hat{T}_0\, \exp\{\int_a^\tau d\xi\,h(\xi,c)\,\frac{\partial}{\partial c}\}\,g(\tau,c)]\,\times\notag \\ &\times  \exp\{-\int_a^\tau d\xi \, [\hat{T}_0\,\exp\{\int_a^\xi d\zeta \,h(\zeta,c)\,\frac{\partial}{\partial c}\}\,\frac{\partial h(\xi,c)}{\partial c}]\}\frac{\partial}{\partial c}\}\,\times\notag \\ &\times\hat{T}_0\, \exp\{\int_a^t d\tau\,h(\tau,c)\,\frac{\partial}{\partial c}\}\,c
\notag
\end{align}
or
\begin{align}
u(t,c)&=\hat{T}_0\, \exp\{\int_a^t d\tau\, [\hat{T}_0\, \exp\{\int_a^\tau d\xi\,h(\xi,c)\,\frac{\partial}{\partial c}\}\,[f(\tau,c)-h(\tau,c)]]\,\times\notag \\ &\times \exp\{-\int_a^\tau d\xi \, [\hat{T}_0\, \exp\{\int_a^\xi d\zeta \,h(\zeta,c)\,\frac{\partial}{\partial c}\}\,\frac {\partial h(\xi,c)}{\partial c}]\}\frac{\partial}{\partial c}\}\,\times\notag \\ &\times \hat{T}_0\, \exp\{\int_a^t d\tau\,h(\tau,c)\,\frac{\partial}{\partial c}\}\,c,
\label{Asol1}
\end{align}
where we can consider $h(t,c)$ as an arbitrary differentiable function.

If it is known the general solution
\[ z(t,c)=\hat{T}_0\, \exp\{\int_a^t d\tau\,h(\tau,c)\,\frac{\partial}{\partial c}\}\,c, \]
then with help of property (\ref{Homomorphy}) we obtain that
\begin{equation}
u(t,c)=\hat{T}_0\, \exp\{\int_a^t d\tau\,\frac{f(\tau,z(\tau,c))-\frac{\partial z(\tau,c)}{\partial \tau}}{\frac{\partial z(\tau,c)}{\partial c}}\,\frac{\partial}{\partial c}\}\,z(t,c).
\label{Asol2}
\end{equation}
As far as $h(t,c)$ is an arbitrary function, therefore preceding expression is valid for any differentiable function $z(t,c)$. The particular solution of equation (\ref{ode1}) with initial condition $u(t,c)|_{t=a}=c$ is obtained when $z(t,c)|_{t=a}=c$.

In spite of obviousness and simplicity the form (\ref{Asol2}) has important meaning. And first of all it gives some recipes for finding ordinary (standard) solution of the initial equation (\ref{ode1}).

In contrast to all above mentioned operator forms of solution the last described is not algorithmical as long as $z(t,c)$ remains to be
an arbitrary function. We do not analyse completely here all consequences of interpretations of solution in form (\ref{Asol2}). Let restrict ourselves by consideration of some solvability conditions only.

Having noted that if
\begin{equation}
\Phi(t,c)=\frac{f(t,z(t,c))-\frac{\partial z(t,c)}{\partial t}} {\frac {\partial z(t,c)}{\partial c}}
\label{Phi}
\end{equation}
corresponds to right-hand part of solvable first-order ODE, then the general solution of given equation can be obtained too.

If we require $\Phi(t,c)=0$, we need to solve the same given equation. In another cases we will need to solve \emph {different} equation and have chance really to solve it. Main difficulties on such a way arise in the fact that we can not circumscribe all solved DE by single expression, so we have the dilemma: if we choose a certain $z(t,c)$ we are not sure that the equation with corresponding $\Phi(t,c)$ is solvable and vice versa -- we are not sure that the equation (\ref{Phi}) in respect to $z(t,c)$ with some $\Phi(t,c)$ selected in advance is solvable.

The situation becomes simpler if we content ourselves with some \emph {solvability conditions}. It is clear that we are interested to pose the broader solvability conditions.

To formulate an algorithmical solvability conditions we have to fix some finite set of functions $z(t,c)$ -- the set of trial functions. Then by using some criterion we successively examine these function trying to find at least one of them (generally speaking, there are a family of such functions) that satisfy the criterion. If such a function is found, then solution of given equation is obtained almost automatically by substitution of this function to (\ref{Asol2}).

The following important classes of equations that permit obtaining of general solutions as a whole can illustrate this approach.

\emph {Linear equations}. Here $\Phi(t,c)=\alpha(t)c+\beta(t)$. It is obvious that solvability condition in this case can be formulated as follows. If $z(t,c)$ is found to be satisfy the criterion
\begin{equation}
\frac{\partial^2}{\partial c^2} \{\frac{f(t,z(t,c))-\frac{\partial z(t,c)}{\partial t}} {\frac {\partial z(t,c)}{\partial c}}\}=0,
\label{Lin}
\end{equation}
then the given equation can be solved through the function $z(t,c)$ and a linear equation. Since in this case $u(t,c)= A(t)z(t,c)+B(t)$ then here we use linear transformations of variable $u(t,c)$. Such type of solving method is due to Liouville. Relatively recent implementations of this approach to CAS make a good showing in solving many of Kamke examples \cite {ChebKol}.

\emph {Equations with separable variables}. Here $\Phi(t,c)=\alpha(t)\beta(c)$. The solvability condition in this case can be formulated as follows. If $z(t,c)$ is found to be satisfy the criterion
\begin{equation}
\frac{\partial^2}{\partial t \partial c} \{ \ln\{\frac{f(t,z(t,c))-\frac{\partial z(t,c)}{\partial t}} {\frac {\partial z(t,c)}{\partial c}}\}\}=0,
\label{Sep}
\end{equation}
then the given equation can be solved through the function $z(t,c)$ and an equation with separable variables. The suitable functions $\alpha(t)$ and $\beta(c)$ under known $z(t,c)$ are easily obtained from (\ref{Phi}).

Reduction of an equation to the one with separable variables of course is the well-known method, but as a matter of fact for solving non-linear equations it is used mainly heuristically simply by guessing appropriate substitution. In resembling a way L. Chan and E.S. Cheb-Terrab have implemented recently so-colled \emph {hypergeometric solutions} method \cite {ChanCheb} where the sets of functions $z(t,c)$ and substitutions are fixed and only second-order linear (or Riccati) equations are considered. In contrast to their method we here have to fix \emph {only} the set of trial functions $z(t,c)$ so our  approach can handle broader family of equations.

\emph {Bernoulli equations}. Here $\Phi(t,c)=\alpha(t)c^\eta+\beta(t)c$ \,($\eta\neq1$). The solvability condition in this case can be formulated as follows. If $z(t,c)$ is found to be satisfy the criterion
\begin{equation}
\frac{\partial}{\partial c}\{\frac { c\,\frac{\partial}{\partial c}\{\frac{f(t,z(t,c))-\frac{\partial z(t,c)}{\partial t}} {\frac {\partial z(t,c)}{\partial c}}\}-\frac{f(t,z(t,c))-\frac{\partial z(t,c)}{\partial t}} {\frac {\partial z(t,c)}{\partial c}}}{c^\eta}\}=0,
\label{Ber}
\end{equation}
then the given equation can be solved through the function $z(t,c)$ and an Bernoulli equation. The suitable functions $\alpha(t)$ and $\beta(t)$ under known $z(t,c)$ are obtained from (\ref{Phi}).

Despite the fact that transformations here are simple enough we do not know any real solving method based on this scheme.

We can easily find examples of equations that are solved in such a way and are not solved say by Maple.

It is obvious that list (\ref{Lin})-(\ref{Ber}) is far from to be complete, analogous solvability conditions can be obtained from another solved equations. And what is more the solution (\ref{Asol2}) can be iterated what lead to schemes with successive  variables changing (substitutions). In other words we can use combinations of conditions too. The general strategy for such type of algorithms needs further development.

\section{Conclusions}
We presented some ways in solving DE by operator method. Besides first-order DE we can obtain operator solutions for linear and non-linear equations of arbitrary order. The properties of its solutions expect further studies.

In conclusion we believe that we succeeded in demonstration the fact that operator forms of DE solutions can be handled analytically none the worse than ordinary functions. In some cases its transformation properties are more comfortable than, e.g., for some special functions.

\end{document}